\documentclass[a4paper,showkeys,floatfix,aps,pre,preprint,groupedaddress]{revtex4-1}
\usepackage{amsmath,amssymb}
\usepackage{graphics,graphicx}
\usepackage{dcolumn,bm}
\usepackage{psfrag}
\usepackage{color}
\begin{document}

%%%% Article title to be placed here
\title{Opinion dynamics: Public and private}

\author{%%%% Author details
Subhadeep Roy $^{1}$ and Soumyajyoti Biswas$^{2}$}

%%%%%%%%% Insert author address here
\affiliation{$^{1}$ The Institute of Mathematical Sciences, Chennai, Tamil Nadu - 600113, India\\
$^{2}$ Department of Physics, SRM University - AP, Andhra Pradesh - 522502,
India }

%%%% Subject entries to be placed here %%%%
%\subject{xxxxx, xxxxx, xxxx}

%%%% Keyword entries to be placed here %%%%
%\keywords{opinion dynamics, kinetic exchange model, phase transition}

%%%% Insert corresponding author and its email address}
%\corres{Soumyajyoti Biswas\\
%\email{soumyajyoti.b@srmap.edu.in}}

%%%% Abstract text to be placed here %%%%%%%%%%%%
\begin{abstract}
We study here the dynamics of opinion formation in a society where we take into account of 
the internally held beliefs and externally expressed opinions of the individuals, which are
not necessarily the same at all times. While these two components can influence one another, 
their difference, both in dynamics and in the steady state, poses interesting scenarios in
terms of the transition to consensus in the society and characterizations of such consensus.
Here we study this public and private opinion dynamics and the critical behavior of the consensus
forming transitions, using a kinetic exchange model.
\end{abstract}
%%%%%%%%%%%%%%%%%%%%%%%%%%%
\maketitle
%%%%%%%%%% Insert the texts which can accomdate on firstpage in the tag "fmtext" %%%%%

%\begin{fmtext}

\section{Introduction}
%%%% Insert A head here

{\it '' But If Everybody's Watching, You Know, All Of The Back Room
Discussions And The Deals, You Know, Then People Get A Little Nervous, To
Say The Least. So, You Need Both A Public And A Private Position." 

-Hillary Clinton }

\vskip0.5cm

It is common, not only among politicians but also for everyone else, to have an internally held belief and an 
externally expressed opinion, which are not always exactly the same. 
However, it is the latter that is revealed in opinion surveys and can be used as a proxy to 
gauge public perception about any issue. Therefore, a difference between the two 
can lead to a deceptive representation of public attitude. Such a scenario was
indeed predicted back in 1981 \cite{king}.
In 2004/2005 New Europe Barometer (NEB) survey in 13 countries showed that over half the population
expressed fear in revealing their opinion \cite{rose}. 
%\end{fmtext}

%%%%%%%%%%%%%%% End of first page %%%%%%%%%%%%%%%%%%%%%
More specific examples of the discrepancies 
between public and private opinions include the attitude towards immigrants, both in the USA \cite{fuss}
and in Europe \cite{ent}, which show an overall increase in the populist anti-immigrant viewes, whicle 
people's perception towards changed very little.  
Peer pressure, 
unsubstantiated basis etc., could be some of the causes for such a difference, wwhich can lead to 
the so-called spiral of silence \cite{silence}.
 
Nevertheless, 
it is also expected that the public and private opinions can also influence one another -- if the externally 
expressed opinion becomes far removed from a firmly held internal belief, then it could be 
unsustainable in the long term for most people. On the other hand, while the difference exists,
any conducted survey would necessarily predict a biased outcome, leading to outright wrong 
perceptions of the public opinion. It is interesting, therefore, to study some of these effects
in a model.

Expressing opinions through numbers is a well established practice \cite{rmp,galam_book,socio_book}. Especially when the
opinion values can take only binary outcomes -- yes/no voting (e.g., Brexit \cite{brexit}), two party 
elections etc., it can be expressed by just $\pm 1$ and 0 representing the neutral population. 
Opinion values can also be continuous within a range $(-1,+1)$, representing the strength of the bias
towards two opposing ideologies.

The evolution of the opinion of any individual can happen through interactions or 'exchanges' with others. 
Given the well-connected nature of the social contacts, such exchanges can in principle happen with any other
individual in the society. This lead to a wide range of studies involving what is called the 'kinetic exchange
models' of opinion formation (see e.g., \cite{lccc,BCS,nuno1,nuno2}). Inspired by the wealth exchange model of similar nature, the opinion
formation model does not include any conservation of the total opinion, unlike the total wealth. Also unlike
the wealth exchange model, it can show a spontaneous symmetry breaking transition, generally known to belong to
the Ising universality class \cite{op_ising}.
This can, however, change due to the topology, the states of the opinions considered and the number of agents taking part in a single interaction \cite{nuno19,sb},
where even a discontinuous transition can be observed. Discontinuous transitions can also be seen in $q$-voter models studied in various topologoies, inertial effects and with
anti-conformist agents \cite{vot1,vot2,vot3,vot4,vot5,vot6}.
 
Other than the phase transition in the steady states, the dynamics of 
such exchange models in general, have been widely studied -- within a bounded confidence limit \cite{Deff,HK,toscani}, with 
multiple types of individuals \cite{galam86,galam2002}, showing coarsening and/or fragmentation of opinions. 
Indeed, the difference between public and private opinions were also studied in agent based models \cite{plos}, but with two
different groups of individuals, classified according to their attitudes towards expressing public views. 

Here we study an opinion dynamics model, where every individual ($i$) has their public ($o_i(t)$) and private ($P_i(t)$) 
opinion value at any time $t$, representing what they express externally and their internal belief, respectively. 
While the public opinion is subjected to the process of kinetic exchange i.e., influencing others or getting influenced 
by the others, the privately held opinion is only subjected to one's own conflict resolutions. These two components evolve
with a coupled dynamics. In what follows, we define the co-evolution dynamics of these components in the model, their transitions
to consensus and the corresponding exponent values, and the influences on the outcomes of any intermediate opinion survey.

\section{Model}
We follow here a kinetic exchange model of opinion, but it has a public and a private component. The public opinion
values evolve following the kinetic exchange rule:
\begin{equation}
o_i(t+1)=o_i(t)+\mu_{ij}o_j(t),
\end{equation}
where $\mu_{ij}=-1$ with probability $p$ and $+1$ otherwise and are annealed i.e., not fixed in time. The individual opinion values are bounded between the extreme values $\pm 1$
i.e., if the equation prescribes $o_i(t+1)$ to be higher than +1 or lower than -1, then it is just kept fixed at -1 or +1, respectively. 

As for the private opinion values, it is expected that a change in the public opinion is necessarily due to a limited conviction on the
earlier opinion, therefore a part of the change will also influence the private opinion:
\begin{equation}
P_i(t+1)=P_i(t)+k(o_i(t+1)-o_i(t)),
\end{equation}
where $k$ is a parameter. Finally, if the difference between the public and private opinion values for a given individual is too high, 
then it can become unsustainable. In that case, the sign of the private opinion is assigned as the public opinion value:
\begin{equation}
o_i(t)=Sgn(P_i(t)) ~~~ \mbox{if} |P_i(t)-o_i(t)|>\delta_i,
\end{equation}
where $\delta_i$ is a tolerance parameter for an individual. 
The selections of the two agents for an interaction ($i$ and $j$) are random i.e., any two agents can interact at a given time. 
While it is known that the number active social contacts for an individual is between 100-200 (Dunbar number \cite{dunbar}) and that
this number is also obtained even in social media contacts \cite{ves}, the number is large enough to be approximated by the mean-field 
interaction assumed here. 
   
For a large value of the tolerance parameter, it is expected that the public opinion dynamics is completely independent of the private opinion
values and therefore should give the known mean-field transition at $p_c=1/4$ \cite{BCS}. The influence, tolerance and conviction parameters
are indeed the relevant variables that are looked at in social sciences, in view of the difference between public and private opinions \cite{scheu}.

The dynamics of the model evolves as follows: Initially, $o_i(t=0)$ are all assigned $\pm 1$ values randomly with equal probability. The initial private opinion
values $P_i(t=0)$, on the other hand, are assigned a continuous value between ($-1,+1$) with uniform probability. Both $o_i(t)$s and $P_i(t)$s are 
bounded within -1 and +1 i.e., if a higher value is obtained by the evolution equations described above, the values are limited to the extreme values.

A single Monte Carlo time step is defined as $N$ updates of two randomly chosen individuals, where $N$ 
is the number of individuals in the society. 

\section{Results}
Here we describe the dynamics and steady state properties of the model with the evolution rules mentioned above and for different values of the 
parameters $p$ (probability of negative public interaction), $k$ (internal influence parameter) and $\delta$ (tolerance parameter). As mentioned above, the model is 
simulated in the mean field limit i.e., a fully connected graph with $N$ individuals. For simplicity, we keep $\delta_i=\delta$ i.e., the tolerance parameter
is the same for all individuals.

\subsection{Phase diagrams}
An important quantity to measure is the average opinion value of the entire society, which indicates the formation or the lack of consensus in the society. 
In this case, of course, there are two such measures -- in the private and in the public opinions. We argue that while the private opinion values matter in the case
of a  voting through secret ballots, an opinion survey would only reveal the public opinion. Therefore, any difference in the values of these two quantities, either during the dynamics or
in the steady state, would lead to a misleading interpretation of the public perception about an issue. 

The average opinion value of the public opinion, which is also generally used as the order parameter for the transition towards consensus, is given by
\begin{equation}
O(t)=\frac{1}{N}|\sum\limits_i o_i(t)|,
\label{op1}
\end{equation}
and similarly for the private opinions, one can define
\begin{equation}
Q(t)=\frac{1}{N}|\sum\limits_i P_i(t)|.
\label{op2}
\end{equation}

\begin{figure*}[tbh]
\centering\includegraphics[width=5.5in]{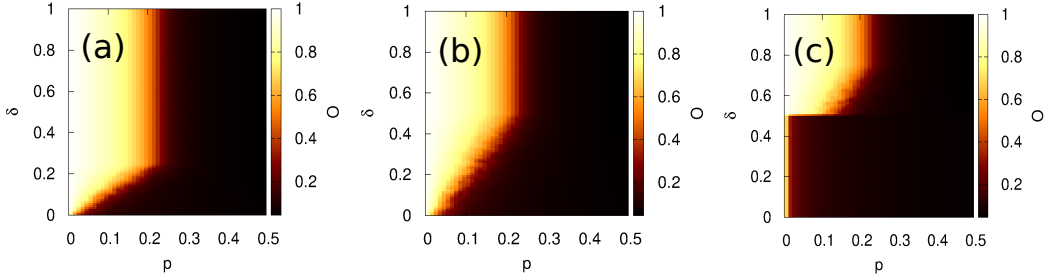}
%%% where xxxxxx name represents "figurename.eps"
\caption{The steady state value of the average public opinion is shown in the $p$ - $\delta$ plane for three different values of $k$, $k=0.5$ in (a), $1.0$ in (b) 
and $1.5$ in (c). High values of $p$ always lead to disordered state. For high values of the tolerance parameter $\delta$, the critical point is the usual $p_c=1/4$,
but for lower $\delta$ values disorder sets in even earlier. For higher values of $k$, the disorder in the private opinion values prevail for low $\delta$, but for
high $\delta$, again an order-disorder transition can be seen.}
\label{fig_pd_pub}
\end{figure*}

In Fig. \ref{fig_pd_pub}, we show the phase diagram of the model in terms of the public opinion values mentioned in Eq. (\ref{op1}). There are several features to be noted. First, for high values of $\delta$, the public opinion dynamics essentially is decoupled from the private one. This is due to a high tolerance of the individuals for the
difference between their public and private view. In this case, a critical point at $p_c=1/4$ is retrieved, as is analytically known for the original version of 
the model \cite{BCS}. However, for a smaller $\delta$, disorder sets in for lower $p$, due to the influence of the private opinion. For high values of $k$, the influence of
the private opinion is so high that for any non-zero $p$ disorder sets in. However, even in this case, when $\delta$ is large, the usual order-disorder transition 
is retrieved.  

\begin{figure*}[tbh]
\centering\includegraphics[width=5.5in]{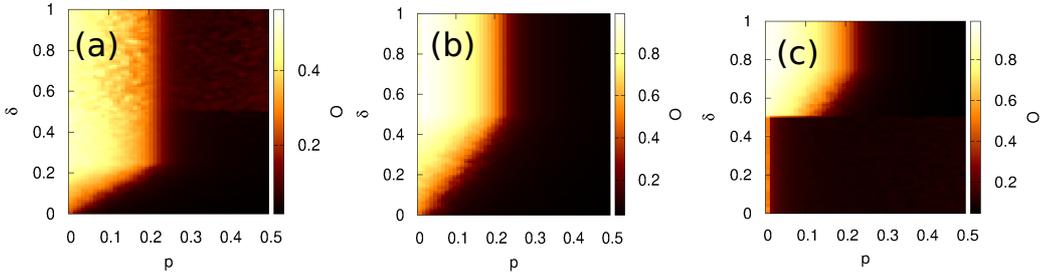}
%%% where xxxxxx name represents "figurename.eps"
\caption{The steady state value of the average private opinion is shown in the $p$ - $\delta$ plane for three different values of $k$, $k=0.5$ in (a), $1.0$ in (b) 
and $1.5$ in (c). In comparing with Fig. \ref{fig_pd_pub} it is seen that the ordering is in the same ranges of the parameters for both the public and the private
opinion values. This means that in the steady state, there is no discrepancy between an election and an opinion survey.}
\label{fig_pd_pvt}
\end{figure*}

\begin{figure*}[tbh]
\centering\includegraphics[width=5.5in]{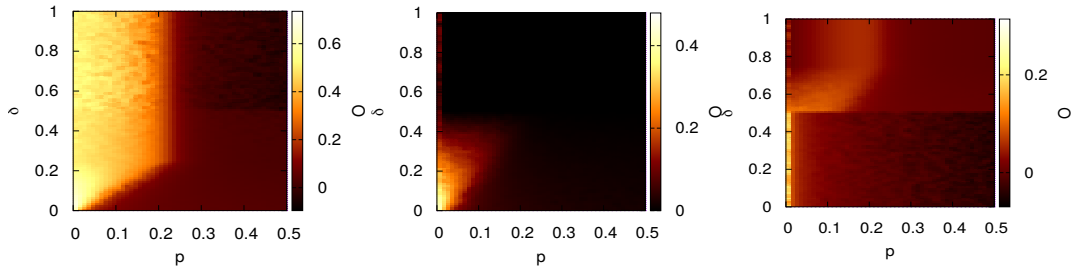}
%%% where xxxxxx name represents "figurename.eps"
\caption{The steady state value of the average private opinion is shown in the $p$ - $\delta$ plane for three different values of $k$, $k=0.5$ in (a), $1.0$ in (b) 
and $1.5$ in (c). In comparing with Fig. \ref{fig_pd_pub} it is seen that the ordering is in the same ranges of the parameters for both the public and the private
opinion values. This means that in the steady state, there is no discrepancy between an election and an opinion survey.}
\label{fig_diff}
\end{figure*}

Fig. \ref{fig_pd_pvt} depicts the phase diagram in terms of the average private opinion described in Eq. (\ref{op2}).  It is seen that in the steady state, the public and private
opinion values show consensus in the same regions of the phase diagram. However, to check if there is any difference in the public and private opinion values in the steady state, 
in Fig. \ref{fig_diff} we plot the difference values in the $p$ - $\delta$ plane. It is seen that a difference exists between the two measures. This is significant, because
it indicates a discrepancy between an election result, where the private opinion is reflected, and 
an opinion survey, where mostly the public opinion is reflected. Indeed, a small variation can lead to a completely opposite results, due to the coarse-graining effects
that are sometimes accompanied with these binary type voting processes (see for example \cite{bcs_us}). 

\subsection{Critical exponents and finite size scaling}
As mentioned before, the transition seen for the high values of $\delta$ (the vertical line in the phase boundary) is the already known in the original version of the model. 
We need to check the transition seen under the influence of smaller $\delta$ values.

Other than the order parameter, its fluctuations near the critical point also reveals the 
associated critical exponent values. Particularly, the Binder cumulant [..], defined as
$U=1-\langle O^2\rangle/3\langle O^2\rangle^2$, and the `susceptibility' 
$V=N\left[\langle O^2\rangle-\langle O\rangle^2\right]$ are useful quantities to look for
finite size scaling. The angular brackets denote configuration average. 
The finite size scaling form for the Binder cumulant is $U\sim f_1\left[(p-p_c)N^{1/\nu}\right]$,
and that for the `susceptibility' is $V\sim N^{\gamma/\nu}f_2\left[(p-p_c)N^{1/\nu}\right]$ and
finally for the order parameter: $O\sim N^{-\beta/\nu}f_3\left[(p-p_c)N^{1/\nu}\right]$, where
$\beta$, $\nu$, $\gamma$ are critical exponents. The advantage of the Binder cumulant is that
all the curve for different values of $N$ cross thrugh the point $p=p_c$, giving a chance to
determine the critical point numerically accurately. This value of the critical point can then be used
for the subsequent finite size scalings mentioned above. 

\begin{figure*}[tbh]
\centering\includegraphics[width=2.5in]{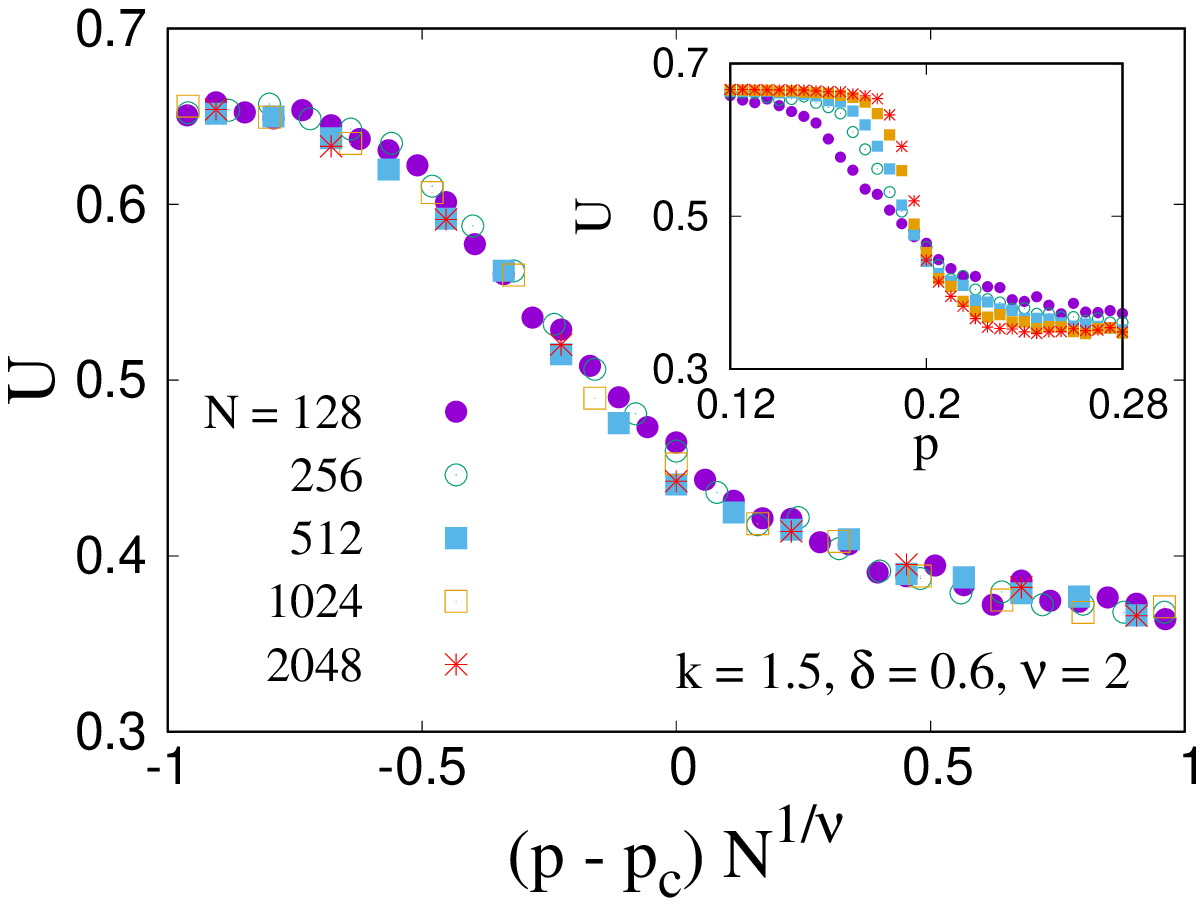}
\centering\includegraphics[width=2.5in]{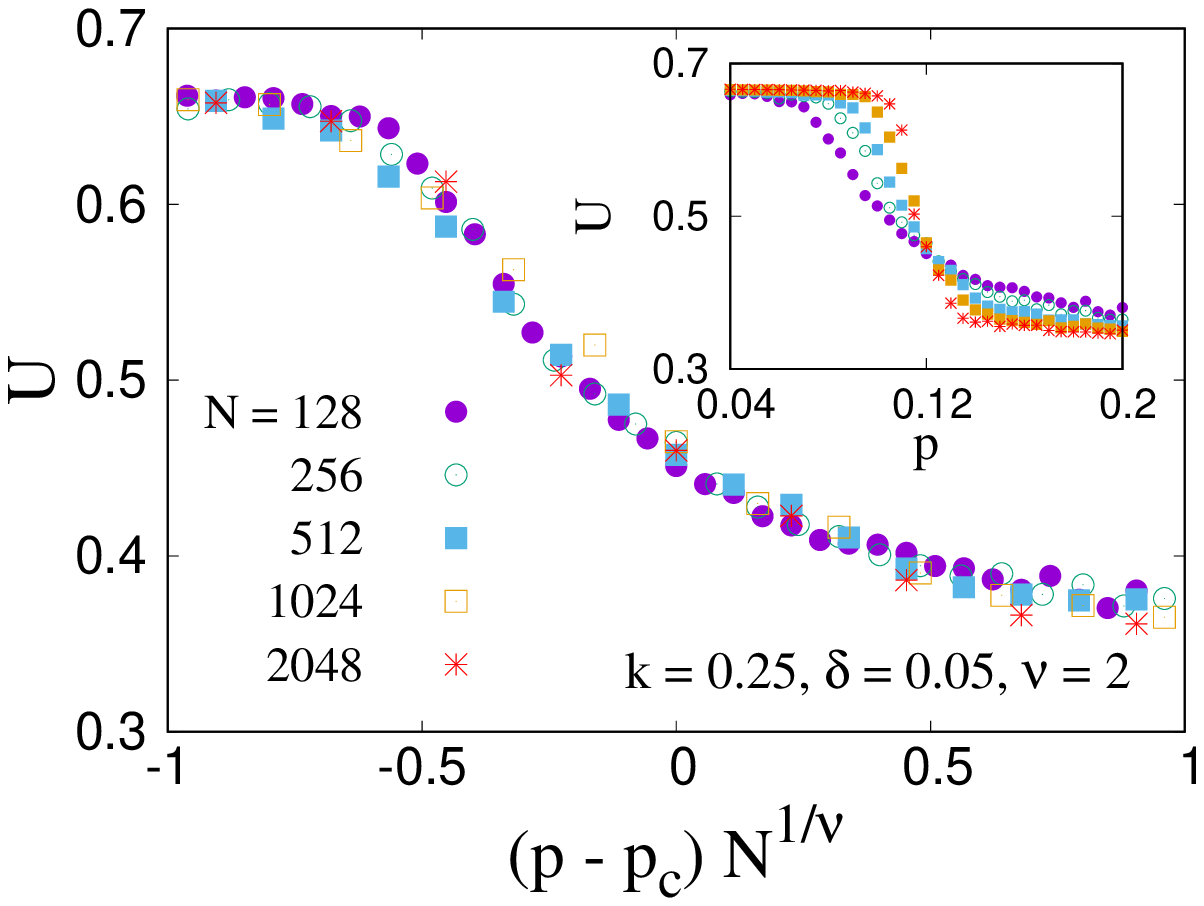}
%%% where xxxxxx name represents "figurename.eps"

\caption{The variations of the Binder cumulant for different system sizes are shown for 
$k=0.25$, $\delta=0.05$ (giving $p_c=0.121\pm 0.002$) and $k=1.5$, $\delta=0.6$ (giving $p_c=0.195\pm 0.002$).
Both the cases, the curves collapse for $\nu=2$.}
\label{binder_cumul}
\end{figure*}
  
In Fig. \ref{binder_cumul}, the Binder cumulants for different system sizes are shown for two sets of ($\delta,k$) values. 
The crossing points of the Binder cumulants give the critical point and the finite size scaling exponent $\nu=2$ is obtained from
the collapse.

\begin{figure*}[tbh]
\centering\includegraphics[width=2.5in]{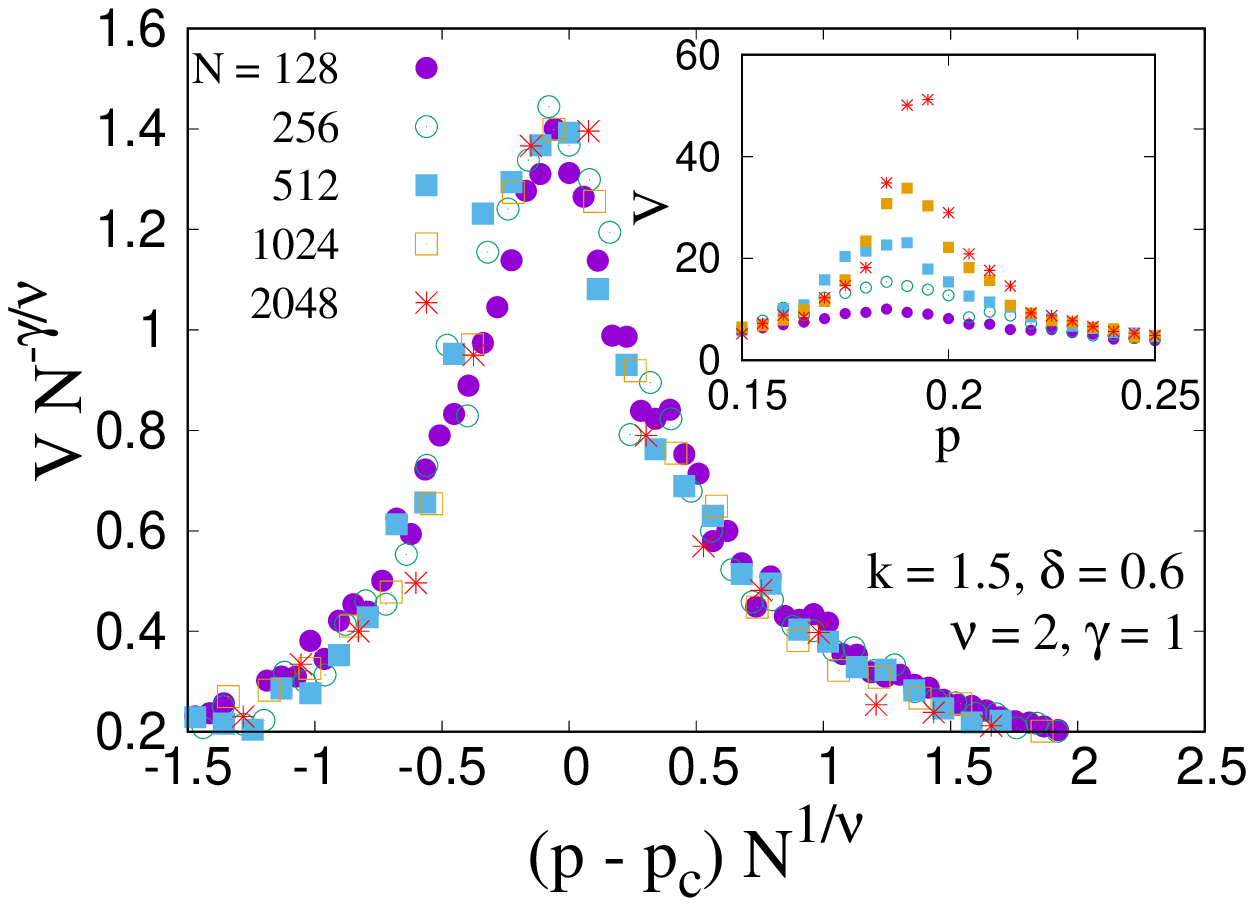}
\centering\includegraphics[width=2.5in]{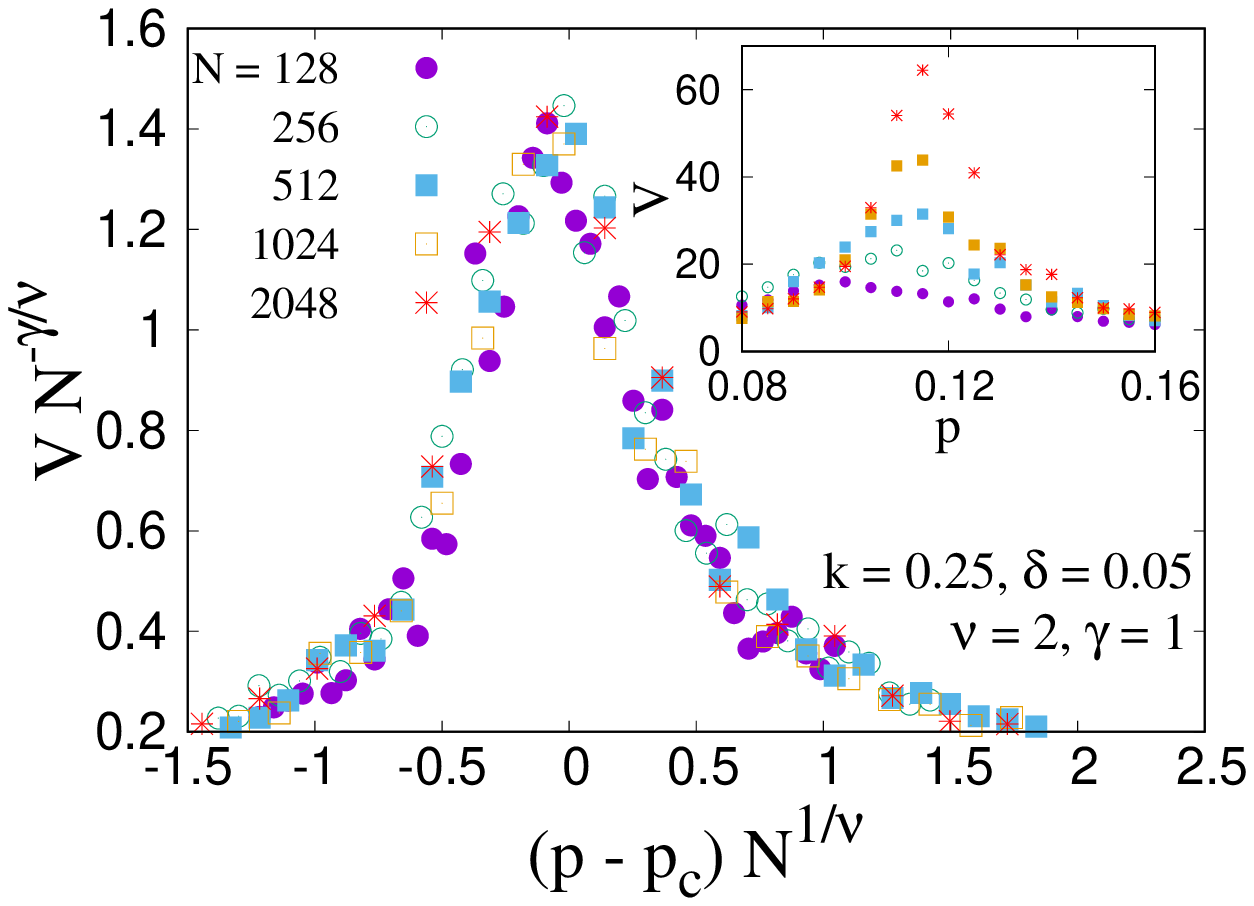}
%%% where xxxxxx name represents "figurename.eps"
\caption{The variations of the susceptibility for different system sizes are shown for 
$k=0.25$, $\delta=0.05$ and $k=1.5$, $\delta=0.6$. Using the critical points obtained from Fig. \ref{binder_cumul},
the finite size scaling analysis is done, which gives $\gamma\approx 1$ (using $\nu=2$).}
\label{suscep}
\end{figure*}

In Fig. \ref{suscep}, the finite size scaling of the susceptibilities are shown. Using the critical points obtained from Fig. \ref{binder_cumul},
the scaling exponent $\gamma=1$ is seen.

\begin{figure}[!h]
\centering\includegraphics[width=2.5in]{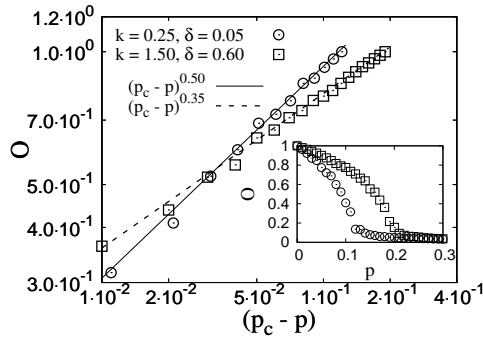}
%%% where xxxxxx name represents "figurename.eps"
\caption{The variation of the order parameter is shown for two combinations of the $k$ and $\delta$ values ($N=2048$). The exponent 
value seems to depend on the parameter set for low values of $k$. For high values, the mean field exponent is retrieved.}
\label{pub_op}
\end{figure}

 Fig. \ref{pub_op} depicts the 
variation of the order parameter (Eq. (\ref{op1})) for two sets of $k$, $\delta$ values. For $k=0.25$ and $\delta=0.05$, the transition with respect to $p$ shows an 
exponent which is different from the usual mean field value. However, for $k=1.50$ an $\delta=0.60$, the mean field exponent value is seen. 

\begin{figure}[!h]
\centering\includegraphics[width=2.5in]{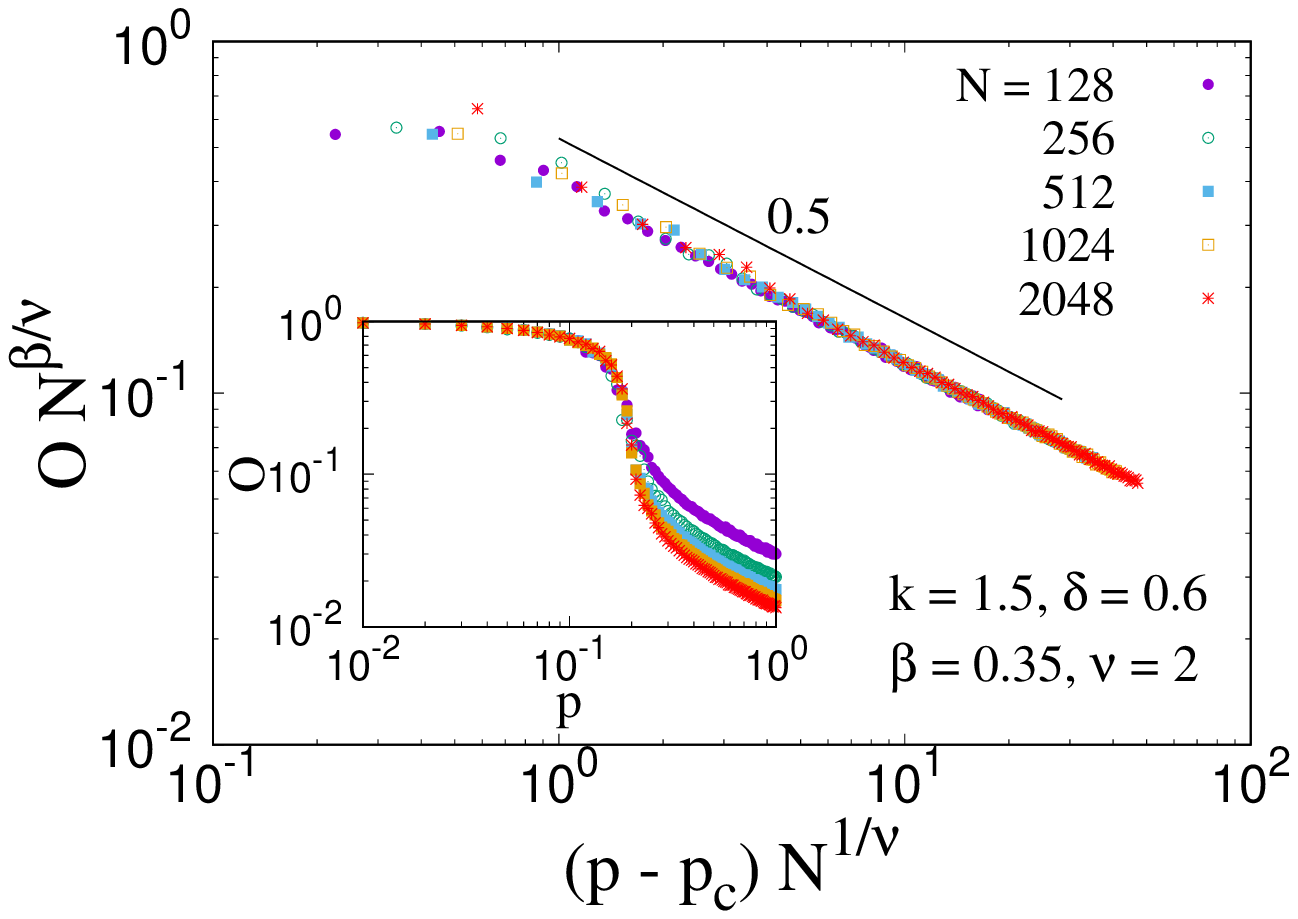}
\centering\includegraphics[width=2.5in]{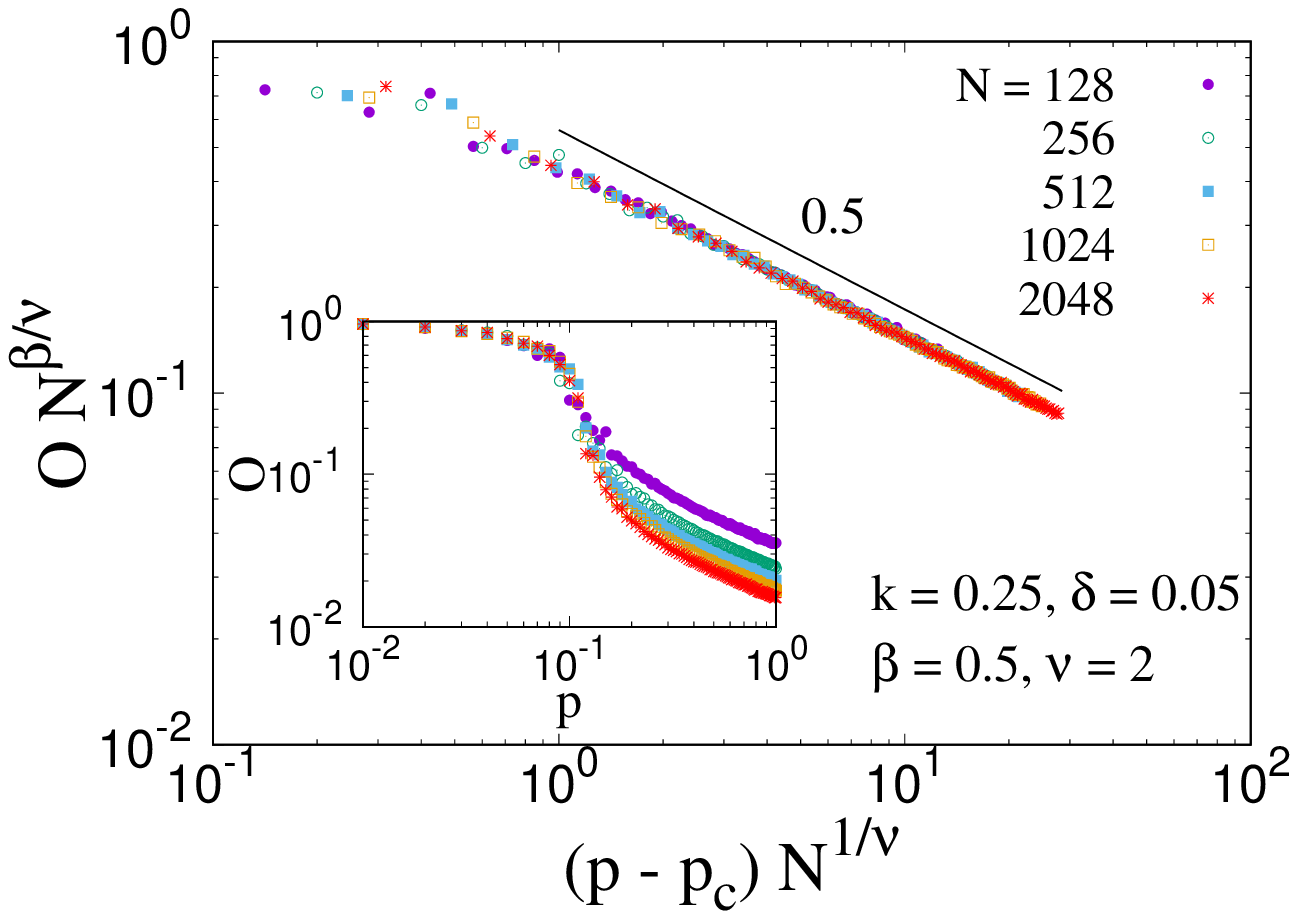}
%%% where xxxxxx name represents "figurename.eps"
\caption{The finite size scaling forms are shown for two different sets of $k$, $\delta$ values. 
With the scaling form taken from Eq. (\ref{fss}), the exponent values for the set $k=1.5$ and $\delta=0.6$
are $\beta=0.35\pm 0.02$ and $\nu=2.00 \pm 0.02$. For the set $k=0.25$, $\delta=0.05$, the exponent values
are $\beta=0.50 \pm 0.02$ and $\nu=2.00 \pm 0.02$.}
\label{fss_op}
\end{figure}

We then do the finite size scaling analysis for these two sets. We assume a finite size scaling of the form:
\begin{equation}
O\sim N^{-\beta/\nu}F \left((p-p_c)N^{1/\nu}\right),
\label{fss}
\end{equation}
where $\nu$ is the effective correlation length exponent. The finite size scaling analysis are shown in Fig. \ref{fss_op}. For the higher value
of $\delta$, the exponents seem to differ from the mean field values. Otherwise, the mean field critical exponents are retrieved. In these cases also, we use the
critical points obtained from the crossing points of the Binder cumulant in Fig. \ref{binder_cumul}. It is important to note here that the exponent values, obtained through
finite size scaling analysis, remain unchanged when calculated from the order parameter of average private opinion ($Q$). 

\begin{figure}[tbh]
\centering\includegraphics[width=2.5in]{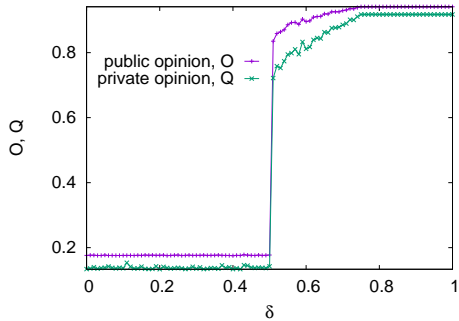}
%%% where xxxxxx name represents "figurename.eps"
\caption{The variations of $O$ and $Q$ with $\delta$ for a fixed value of $p=0.05$ and $k=1.5$. Both the quantities show a discontinuous jump, suggesting a first order transition along this line.}
\label{jump}
\end{figure}

\begin{figure*}[tbh]
\centering\includegraphics[width=5.5in]{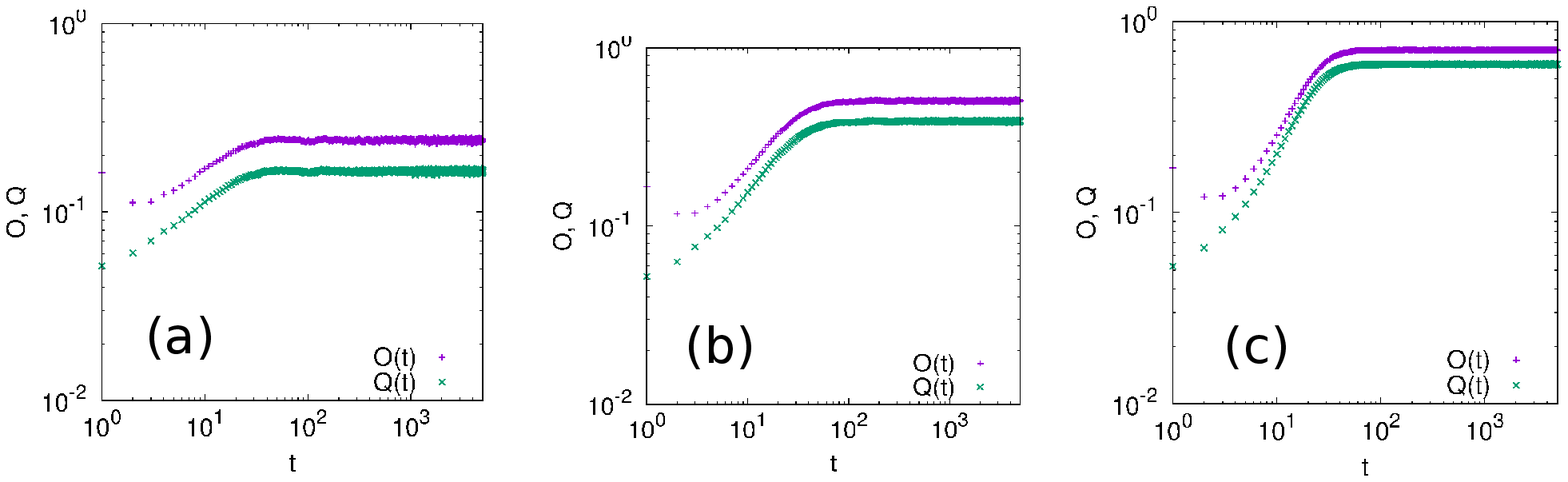}
%%% where xxxxxx name represents "figurename.eps"
\caption{The variations of $O(t)$ and $Q(t)$ with $t$, for three sets of parameter values (a) ($\delta=0.1$, $k=1.0$, $p=0.1$),
(b) ($\delta=0.2$, $k=1.0$, $p=0.1$) and (c) ($\delta=0.3$, $k=1.0$, $p=0.1$). In all cases, the difference between
$O(t)$ and $Q(t)$ persists throughout the dynamics.}
\label{dyn}
\end{figure*}

We also studied the transition with respect to $\delta$, for a fixed value of $p$ for $k=1.5$. The average values of both the public and private opinion values show a
discontinuous jump, suggesting a first order transition along this line (Fig. \ref{jump}).

Finally, we look at the dynamics of $O(t)$ and $Q(t)$ to see if the difference noted in Fig. \ref{fig_diff} also can be seen during the dynamics. In Fig. \ref{dyn},
we show this for three sets of parameter values ($\delta=0.1$, $k=1.0$, $p=0.1$), ($\delta=0.2$, $k=1.0$, $p=0.1$) and ($\delta=0.3$, $k=1.0$, $p=0.1$). In all cases, we see 
that the difference persists during the entire dynamics of the model. This again implies that even at any intermediate time, a survey and an election result might differ, no matter how good the survey statistics are.

\vspace*{-5pt}

\section{Conclusion}
The differences between the public position and privately help belief is a common feature in human behavior \cite{king,rose}. In political scenarios, involving binary choice elections and their preceding opinion surveys, such difference can plan a crucial role in determining the accuracy and subsequent credibility of such surveys.
Indeed, it is known that common perceptions of the attitude of the general public are known to differ from their individual beliefs \cite{fuss,ent}. 
 In this work, we attempted to model the dynamics of those two components of the opinion of the individuals in a society through a simple kinetic exchange model. The public and private opinion values co-evolve in a couple manner, influencing one another during the course of dynamics. If the difference between the two is too high, given by a tolerance factor, then the private opinion dominates. 

We show, however, that even though a consensus is spontaneously reached in the society through the such dynamics for both these components in the same part of the phase diagram, these two components continue to differ statistically significantly throughout the dynamics and in the steady state of the model (see Figs. \ref{fig_diff}, \ref{dyn}). Such a difference is crucial, especially in closely fought elections, in terms of predicting the outcome of such an election through an opinion survey. 

The order of the phase transition to consensus depends on the parameter $k$ (coupling between the private and public opinion values). 
While the kinetic exchange opinion models are known to show Ising universality \cite{op_ising}, in the contineous phase transition for this model, 
the exponent value of the order parameter differs from the Ising universality class for larger values
of the tolerence parameter $\delta$. 

In conclusion, the co-evolution of publicly expressed and privately held opinions of interacting agents in a society can produce a difference between these two quantities, which is known to exist in society. This can shed light on the differences observed in opinion surveys and election results, and further studies regarding
the co-evolution dynamics and their possible hysteresis behaviour near the discontinuous transition could be useful in explaining the 
opposite behavior of the public perception and election outcomes. 

\vskip6pt

\enlargethispage{20pt}

%\ethics{Insert ethics statement here if applicable.}

%\dataccess{All data generated through simulations of the model described in the paper. The code can be made available on request.}

%\aucontribute{ SB designed the model. SR carried out the simulations. SB wrote the manuscript. Both authors have read and approved the manuscript.}

%\competing{The author(s) declare that they have no competing interests.}

%\funding{Insert funding text here.}

%\ack{Insert acknowledgment text here.}

%\disclaimer{Insert disclaimer text here if applicable.}

%%%%%%%%%% Insert bibliography here %%%%%%%%%%%%%%

\end{document}